\documentclass[twocolumn,aps,amsmath,showpacs,amssymb,floatfix]{revtex4}
\usepackage{tabularx}
\usepackage{bm}
\usepackage{subfigure}
\usepackage{euscript}
\usepackage{graphicx}
\usepackage{color}
\usepackage[colorlinks=true,linkcolor=blue]{hyperref}%
\usepackage{amsfonts}
\usepackage{exscale}
\usepackage{amsbsy}

\begin{document}

\title{Magnetic Impurities in Two-Dimensional Superfluid Fermi Gas\\ with Spin-Orbit Coupling}

\author{Zhongbo Yan}
\author{Xiaosen Yang}
\author{Liang Sun}
\author{Shaolong Wan}
\email[]{slwan@ustc.edu.cn} \affiliation{Institute for Theoretical
Physics and Department of Modern Physics University of Science and
Technology of China, Hefei, 230026, \textbf{P. R. China}}
\date{\today}

\begin{abstract}
We consider magnetic impurities in a two dimensional superfluid
Fermi gas in the presence of spin-orbit coupling. By using the
methods of t-matrix and Green's function, we find spin-orbit
coupling has some dramatic impacts on the effects of magnetic
impurities. For the single impurity problem, the number of bound
states localized around the magnetic impurity is doubled. For the
finite concentration $n$ of impurities, the energy gap is reduced
and the density of states in the gapless region is greatly
modified.

\end{abstract}


\maketitle

\section{Introduction}

In the last few years, topological insulator (TI) and topological
superconductor (TSC) \cite{M. Z. Hasan, X. L. Qi} have arisen a
lasting interest in condensed matter systems. In such systems,
spin-orbit coupling plays a fundamental role to make the TI and
TSC different from the traditional insulator and traditional BCS
superconductor significantly. In cold atomic field, the superfluid
of Fermi gases have been realized in ${^{40}}K$ \cite{C. A. Regal}
and ${^{6}}Li$ \cite{M. Bartenstein, M. W. Zwierlein, J. Kinast},
furthermore, spin-orbit coupling has also been realized  by the
more recent experimental achievement of synthetic gauge field
\cite{Y. -J. Lin1, Y. -J. Lin2}. By Introducing the effect of
spin-orbit coupling, new phases analog to TSC and new phenomena
may emerge in cold superfluid Fermi gases, e.g. topological
superfluid and topological phase-separation \cite{Xiaosen Yang}.
Since a cold atomic system is more adjustable, it can provide a
more controllable way to study such phases under different
conditions, e.g. magnetic impurities.

From TI and TSC, we know that spin-orbit coupling combining with
Zeeman field greatly change the energy spectrum and make
topological phases emerge. Therefore, to figure out how the
spin-orbit coupling affects the effect of magnetic impurity, which
likes a local magnetic moment, is a very interesting problem.
Recently, people have paid attention to studying on the Kondo
effect in the presence of spin-orbit coupling \cite{Rok Zitko,
Mahdi Zarea, L. Isaev}, however, in this paper, we will study the
effects of classical magnetic impurity in superfluid Fermi gases
in the presence of spin-orbit coupling. Since cold Fermi gas
system is much cleaner than metallic superconductor and the
strength of pairing interaction can be tuned by adjusting the
threshold energy of Feshbach resonance, obviously, it's a better
platform to detect the effects of dilute magnetic impurities.

The effects of magnetic impurities are particularly important in
the case of superconductor, and so superfluidity. Magnetic
impurities which can flip the spins of electrons interfere with
the pairing and reduce the gap. A sufficient concentration of
magnetic impurities leads to gapless superconductivity, where the
superconductive order parameter still exist but the BCS excitation
gap is absent \cite{A. A. Abrikosov, K. Maki}. Such a
pairing-breaking effect is absent in the case of nonmagnetic
impurities (known as Anderson's theorem), so in the following, we
only consider the effects of magnetic impurities.

In this paper, we discuss the effects of magnetic impurities in a
two dimensional superfluid Fermi gas in the presence of spin-orbit
coupling. So far, nonmagnetic impurities have been realized in
Bose gases by using other species of atoms \cite{C. Zipkes} and
laser light \cite{G. Modugno}. Recently, a single
spin-$\downarrow$ placed in a Fermi sea of spin-$\uparrow$ atoms
is observed to behave as a mobile impurity to form Fermi polaron
\cite{A. Schirotzek, S. Nascimbene}. To realize magnetic
impurities in a superfluid Fermi gas, Y. Ohashi \cite{Y. Ohashi}
suggests and confirms that nonmagnetic impurities accompanied by
localized spin-$\uparrow$ can be viewed as magnetic impurities.
Based on these, we believe it is not far away to detect the
effects of magnetic impurities in a superfluid Fermi gas with
spin-orbit coupling in real experiments.

The paper is organized as follows. In Sec.\ref{sec2}, we consider
the case of single magnetic impurity and calculate the energy of
bound state in the presence of spin-orbit coupling. In
Sec.\ref{sec3}, we generalize the single impurity case to the one
of concentration $n$. In Sec.\ref{sec4}, a conclusion is given.

\section{Single Impurity Model}
\label{sec2}

Now, we consider a two dimensional ultra-cold atomic system with
single magnetic impurity and spin-orbit coupling, which is
described by
\begin{eqnarray}
H&=&H_{0}+H_{imp}, \nonumber \\
H_{0}&=&\sum_{k\sigma}\varepsilon_{k}c_{k\sigma}^{\dag}c_{k\sigma}+
\sum_{k}\alpha_{R}k
\left(e^{i\varphi_{k}}c_{k\uparrow}^{\dag}c_{k\downarrow}+
e^{-i\varphi_{k}}c_{k\downarrow}^{\dag}c_{k\uparrow} \right) \nonumber \\
&& -\Delta\sum_{k}\left(c_{k\uparrow}^{\dag}c_{-k\downarrow}^{\dag}+c_{-k\downarrow}c_{k\uparrow} \right),\nonumber \\
H_{imp}&=&\frac{U}{2}s_{z}\sum_{k,k^{'},\sigma}c_{k\sigma}^{\dag}\sigma
c_{k^{'}\sigma}, \label{1}
\end{eqnarray}
where $\varepsilon_{k}=k^{2}/2m-\mu$, $\alpha_{R}$ is the strength
of spin-orbit coupling, $c_{k\sigma}^{\dag}(c_{k\sigma})$ is
creation(annihilation) operator with pseudo-spin $\sigma$,
$\Delta$ is the order parameter, $U$ is the on-site interaction
strength and $s_{z}=\pm s$ is the number of excess spin-up atoms
localized around impurity (see Refs.\cite{Y. Ohashi}). For
simplicity, here we completely neglect the effect of the local
spatial variation of the order parameter around a paramagnetic
impurity.

In the Nambu spinor presentation $\Psi_{k}^{\dag} =
(c_{k\uparrow}^{\dag}, c_{k\downarrow}^{\dag}, c_{-k\downarrow},
c_{-k\uparrow})$, $H_{0}$ can be rewritten as
\begin{eqnarray}
H_{0}&=&\frac{1}{2}\sum_{k}\Psi_{k}^{\dag}H(k)\Psi_{k}, \nonumber \\
H(k)&=&\varepsilon_{k}\tau_{3}-\Delta\rho_{1}\tau_{3} +\alpha_{R}
k \left(e^{i\varphi_{k}}\tau_{+} +e^{-i\varphi_{k}}\tau_{-}
\right), \label{2}
\end{eqnarray}
where $\rho_{i}$ and $\tau_{i} (i=1,2,3)$ are Pauli matrices in
particle-hole space and spin space, respectively, and
$\tau_{\pm}=(\tau_{1} \pm i\tau_{2})/2$. According to Ref.\cite{H.
Shiba} or the method of the coherent state path integral
\cite{Alexander Altland}, we can have
\begin{eqnarray}
G_{0}(k,\omega)=\frac{1}{\omega-\varepsilon_{k}\tau_{3}+\Delta\rho_{1}\tau_{3}-\alpha_{R}k(e^{i\varphi_{k}}\tau_{+}
+e^{-i\varphi_{k}}\tau_{-})}, \nonumber \\
\label{3}
\end{eqnarray}
$G_{0}(k,\omega)$ is related to the full Green function $G(k,\omega)$ in the following form
\begin{eqnarray}
G(k,k^{'},\omega)=G_{0}(k,\omega)\delta_{kk^{'}}+G_{0}(k,\omega)t(\omega)G_{0}(k^{'},\omega).
\label{4}
\end{eqnarray}
After averaging over the direction of the spin, the non-flip part of the t-matrix, $t(\omega)$, is given by
\begin{eqnarray}
t(\omega)=\frac{(\frac{U}{2}s)^{2}F(\omega)}{1-(\frac{U}{2}s)^{2}F(\omega)^{2}}, \label{5}
\end{eqnarray}
where $F(\omega)$ is
\begin{eqnarray}
F(\omega)=\sum_{k}G_{0}(k,\omega). \label{6}
\end{eqnarray}

We are interested in the localized excited state in the energy gap
$\Delta$, so we must find poles of $t(\omega)$. For
$|\omega|<\Delta$, by integrating out $k$ in Eq.(\ref{6}) (see
details in Appendix), we obtain
\begin{eqnarray}
F(\omega)=-\pi\rho(0)\frac{\omega-\Delta\rho_{1}\tau_{3}-m\alpha_{R}^{2}\rho_{3}}{\sqrt{\Delta^{2}-\omega^{2}}},\label{7}
\end{eqnarray}
where $\rho(0)$ is the density of states at the Fermi level in the
normal state. Based on Eqs.(\ref{5}) and (\ref{7}), the poles of
$t(\omega)$ are given by (see details in Appendix)
\begin{eqnarray}
\omega_{1}&=&\pm \left(1-\frac{1}{8}A\frac{m^{4}\alpha_{R}^{8}}{\Delta^{4}} \right)\Delta, \nonumber \\
\omega_{2}&=&\pm
\left(\frac{1-A}{1+A}-\frac{A}{1+A}\frac{m^{2}\alpha_{R}^{4}}{\Delta^{2}}
\right)\Delta, \label{8}
\end{eqnarray}
where $A=(Us\pi\rho(0)/2)^{2}$. From Eq.(\ref{8}), it's easy to
see, no matter how small $A$ may be, two bound states always
appear, this result is quite different from the case in the
absence of spin-orbit coupling. However, for $\alpha_{R}=0$, there
are only two poles in the energy gap, and the poles reduce to
$\omega=\pm[(1-A)/(1+A)]\Delta$, just the same result as in
Refs.\cite{H. Shiba,Y. Ohashi, Lei Jiang}. Increasing the strength
of spin-orbit coupling, the energy of the low-lying state which is
localized around the impurity goes to zero. From Eq.(\ref{8}), we
can see, if we increase the strength of $U$ and keep $A<1$, the
combined effects of magnetic impurity and spin-orbit coupling is
strengthened, and the energy of the bound state will go to zero
faster than the case in the absence of spin-orbit coupling.

\section{Impurity of Concentration n}
\label{sec3}

Now, we generalize single impurity to the case for finite
concentration n. After averaging over the random distribution of
the impurities, the Green's function recovers the translational
invariance, and the renormalized Green's function is given by
\begin{eqnarray}
G(k,\omega)=\frac{1}{\tilde{\omega}-\varepsilon_{k}\tau_{3}
+\tilde{\Delta}\rho_{1}\tau_{3}-\alpha_{R}k(e^{i\varphi_{k}}\tau_{+}
+e^{-i\varphi_{k}}\tau_{-})}, \nonumber \\
 \label{9}
\end{eqnarray}
where $\tilde{\omega}$ and $\tilde{\Delta}$ are the renormalized
frequency and order parameter, respectively. $\tilde{\omega}$ and
$\tilde{\Delta}$ are determined by Dyson equation
\begin{eqnarray}
G(k,\omega)=G_{0}(k,\omega)+G_{0}(k,\omega)\Sigma(k,\omega)G(k,\omega),\label{10}
\end{eqnarray}
another useful form of this formula is
\begin{eqnarray}
G(k,\omega)^{-1}=G_{0}(k,\omega)^{-1}-\Sigma(k,\omega), \label{11}
\end{eqnarray}
where $G_{0}(k,\omega)$, the Green's function in the absence of impurities, is given by Eq.(\ref{3}).
In Born approximation, the self energy $\Sigma(k,\omega)$ is given by \cite{K. Maki}
\begin{eqnarray}
\Sigma(k,\omega)=-\frac{1}{2\tau_{s}}\frac{\tilde{\omega}-\tilde{\Delta}\rho_{1}\tau_{3}-m\alpha_{R}^{2}\rho_{3}}
{\sqrt{\tilde{\Delta}^{2}-\tilde{\omega}^{2}}}, \label{12}
\end{eqnarray}
where $\tau_{s}$ is the relaxation time and is given by
\begin{eqnarray}
\tau_{s}=\pi\rho(0)ns^{2}U^{2}. \label{13}
\end{eqnarray}
By substituting Eq.(\ref{12}) into Eq.(\ref{11}), we obtain
\begin{eqnarray}
\tilde{\omega}_{1}&=&\omega_{1}+\frac{1}{2\tau_{s}}\frac{\tilde{\omega}_{1}+m\alpha_{R}^{2}}{\sqrt{\tilde{\Delta}_{1}^{2}-\tilde{\omega}_{1}^{2}}}, \nonumber \\
\tilde{\Delta}_{1}&=&\Delta_{1}-\frac{1}{2\tau_{s}}\frac{\tilde{\Delta}_{1}}{\sqrt{\tilde{\Delta}_{1}^{2}-\tilde{\omega}_{1}^{2}}}, \label{14}
\end{eqnarray}
and
\begin{eqnarray}
\tilde{\omega}_{2}&=&\omega_{2}+\frac{1}{2\tau_{s}}\frac{\tilde{\omega}_{2}-m\alpha_{R}^{2}}{\sqrt{\tilde{\Delta}_{2}^{2}-\tilde{\omega}_{2}^{2}}}, \nonumber \\
\tilde{\Delta}_{2}&=&\Delta_{2}-\frac{1}{2\tau_{s}}\frac{\tilde{\Delta}_{2}}{\sqrt{\tilde{\Delta}_{2}^{2}-\tilde{\omega}_{2}^{2}}},
\label{15}
\end{eqnarray}
where $\Delta_{1} = \Delta_{2} = \Delta$. They are quite different
from the case without spin-orbit coupling, as before, with
spin-orbit coupling, the number of poles of t-matrix has increased
from two to four. By introducing a new auxiliary parameter $u_{i}$
defined by $u_{i}\equiv\tilde{\omega}_{i}/\tilde{\Delta}_{i}$
$(i=1,2)$, Eqs.(\ref{14})-(\ref{15}) are reduced as
\begin{eqnarray}
\frac{\omega_{i}}{\Delta_{i}}&=&u_{i} \left(1-\zeta_{i}\frac{1}{\sqrt{1-u_{i}^{2}}} \right)+(-1)^{i}v_{i}\frac{\zeta_{i}}{2\sqrt{1-u_{i}^{2}}-\zeta_{i}}, \nonumber \\
\zeta_{i}&=&\frac{1}{\tau_{s}\Delta_{i}},
~~~~~~v_{i}=\frac{m\alpha_{R}^{2}}{\Delta_{i}}. \label{16}
\end{eqnarray}

\begin{figure}
\subfigure{\includegraphics[width=4cm,height=4cm]{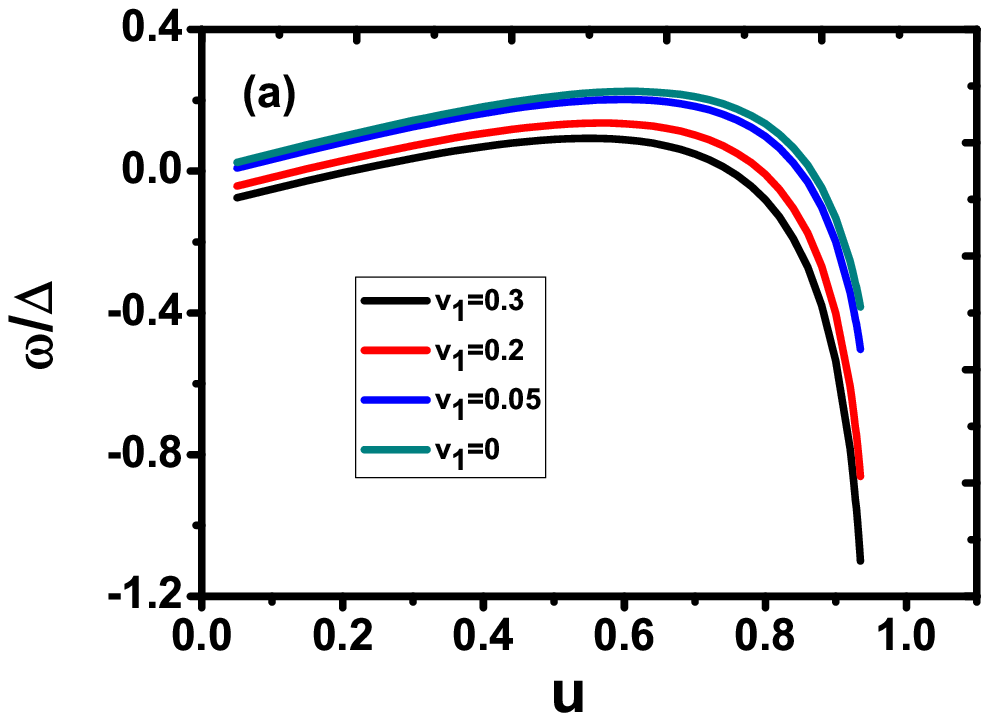}}
\subfigure{\includegraphics[width=4cm,height=4cm]{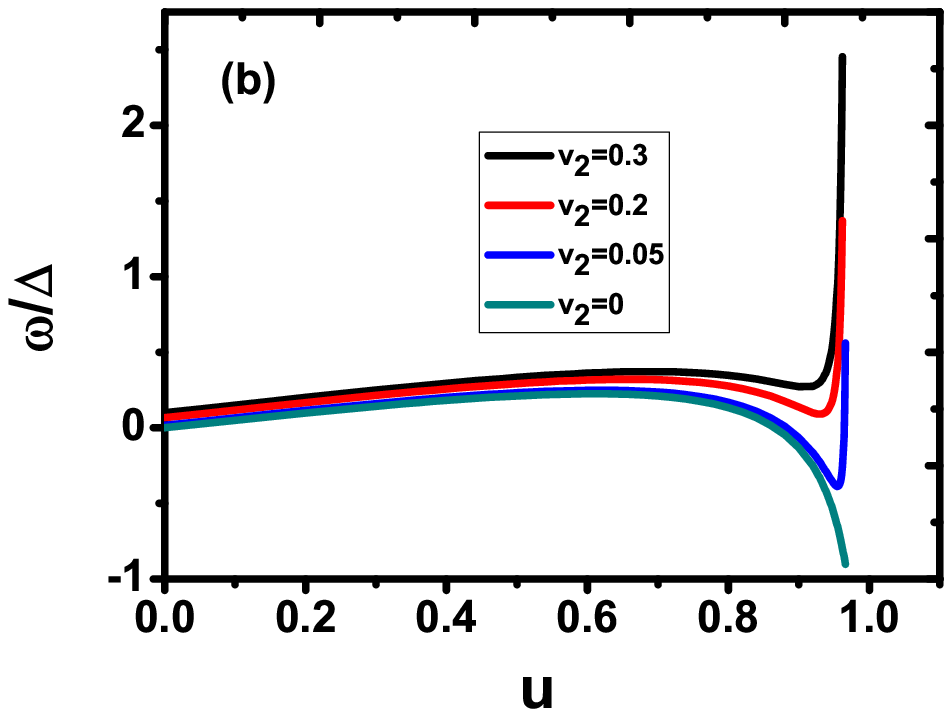}}
\subfigure{\includegraphics[width=4cm,height=4cm]{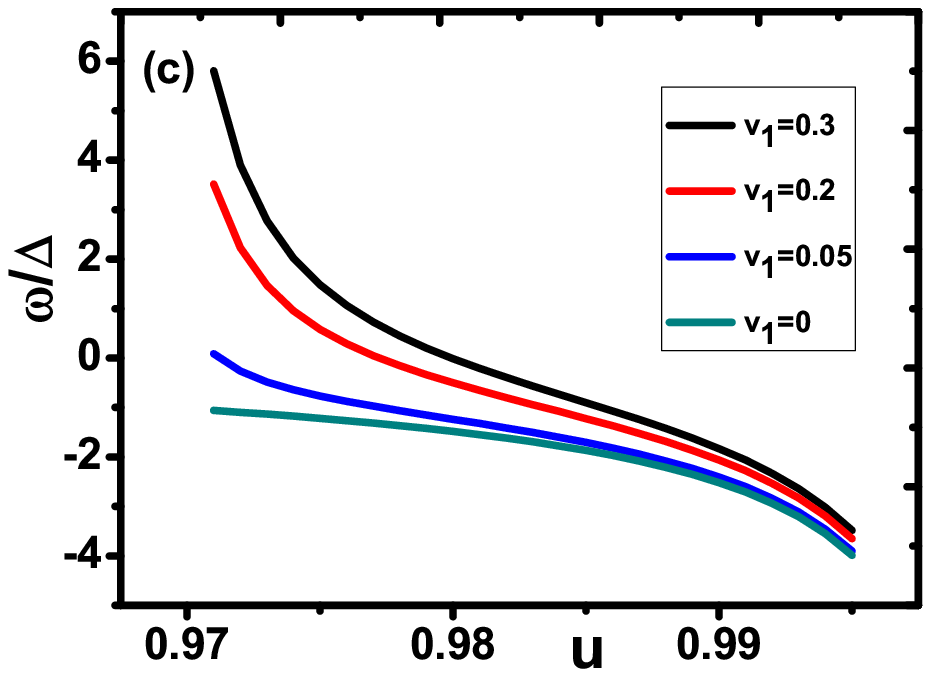}}
\subfigure{\includegraphics[width=4cm,height=4cm]{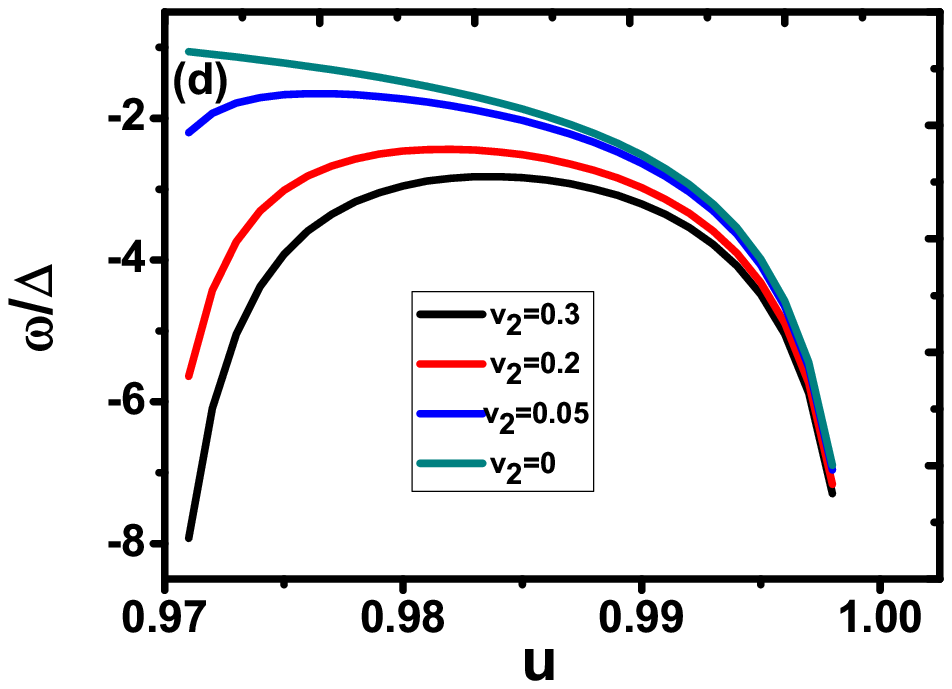}}
\caption{ (Color online) The parameters of Eq.(\ref{16}) are:
$\zeta=0.5$. The values of $v_{i}$ are shown in the figure. (a)
There is a maximum for every fixed $v_{1}$, and the value of the
maximum decreases monotonically with $v_{1}$. (b) There is no
maximum, when $u$ goes to the critical value $u_{c}$, the value of
$\omega/\Delta$ diverges. (c) When $u\in(u_{c},1)$,
$\omega/\Delta$ decreases with $u$ monotonically. (d) There is a
maximum for every fixed non zero $v_{2}$, and the value of the
maximum decreases monotonically with $v_{2}$ } \label{Fig.1}
\end{figure}

Plotting the above equation in the $u \omega$-plane (as shown in
Fig.\ref{Fig.1}), we can see the case in the absence of spin-orbit
coupling is quite different from the one with spin-orbit coupling.

\begin{figure}
\subfigure{\includegraphics[width=9cm,height=6cm]{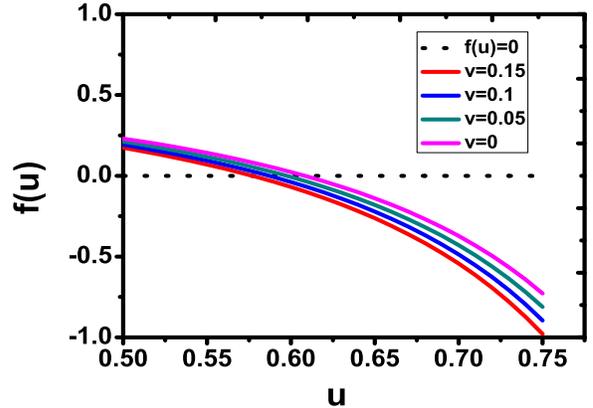}}
\caption{ (Color online) The parameters of Eq.(\ref{17}) are:
$\zeta=0.5$. $f(u)= \frac{1}{\Delta}\frac{\partial\omega}{\partial
u}$. By increasing $v$ from $v=0.05$ to $v=0.15$, $u_{0}$, the $u$
corresponds to the maximum of $\frac{\omega}{\Delta}$, changes
slowly. } \label{Fig.2}
\end{figure}

First, for $\zeta<2$ case, with the spin-orbit coupling, there are
two poles in Eq.(\ref{16}), one is at
$u_{c}=\sqrt{1-\zeta^{2}/4}$, and the other is the original one at
$u_{c}=1$. In the absence of spin-orbit coupling, for $v=0$,
$\omega$ goes through a maximum and afterwards decrease
monotonically, this is also the characteristic of
Fig.\ref{Fig.1}(a)(c)(d). However, there is no such characteristic
in Fig.\ref{Fig.1}(b), after $\omega$ goes through a finite
maximum, $\omega$ will not decreases monotonically, in the
opposite, when $u$ goes to $u_{c}$, $\omega$ is positive and
diverges. So only the maximum value of $\omega$ in
Fig.\ref{Fig.1}(a) has the meaning of the energy gap $\omega_{g}$
(in the following, we only consider this case). It is determined
by the condition
\begin{eqnarray}
\frac{1}{\Delta}\frac{\partial\omega}{\partial u}&=&1-\zeta(1-u^{2})^{-3/2}-\frac{2vu\zeta}{\sqrt{1-u^{2}}(2\sqrt{1-u^{2}}-\zeta)^{2}} \nonumber \\
&=&0. \label{17}
\end{eqnarray}
This equation is difficult to solve. However, by numerical
calculating, we find that if $v$ be sufficiently small, like
$v=0.05$ in Fig.\ref{Fig.1}, we can safely ignore the last term in
Eq.(\ref{17}) (see Fig.\ref{Fig.2}), and obtain

\begin{eqnarray}
u_{0}&\simeq&(1-\zeta^{2/3})^{1/2},  \nonumber \\
\omega_{g}&\simeq&\Delta
\left[(1-\zeta^{2/3})^{3/2}+\frac{v}{1-2\zeta^{-2/3}}\right].
\label{18}
\end{eqnarray}
We see that the solution of Eq.(\ref{17}) exist only when
$\zeta<1$. In fact, even we consider the effect of spin-orbit
coupling and modify $u_{0}$ to $(1-a-\zeta^{2/3})^{1/2}$, to first
order, the expression of $\omega_{g}$ in Eq.(\ref{16}) will not
change. Based on above, we can expand Eq.(\ref{16}) around $u_{0}$
and obtain
\begin{eqnarray}
\frac{\omega-\omega_{g}}{\Delta}=-\frac{3}{2}\zeta^{-\frac{2}{3}}
\left[(1-\zeta^{\frac{2}{3}})^{\frac{1}{2}}
+\frac{v(8-6\zeta^{\frac{2}{3}})}{3(2-\zeta^{\frac{2}{3}})^{3}} \right](u-u_{0})^{2}. \nonumber \\
\label{19}
\end{eqnarray}

Second, for $1<\zeta<2$ case, $\omega$ decreases monotonically as
$u$ increases between $[0,u_{c})$. This corresponds to $\omega_{g}
= 0$.
Because of the presence of spin-orbit coupling, from
Eq.(\ref{18}), we can see that, when $\zeta<1$, $\omega_{g}$ can
reach zero. This indicates that spin-orbit coupling modifies
strongly the critical concentration $n_{c}$ ( $n_{c}$ corresponds
to $\zeta=1$ in the absence of spin-orbit coupling), where the gap
in the energy spectrum $\omega_{g}$ vanishes and the gapless
superfluidity appears.

For small values of $\omega$, an asymptotic expression of $u$ based on Eq.(\ref{15})
is given by
\begin{eqnarray}
u=&&i\sqrt{(\zeta+\alpha)^{2}-1}\nonumber \\
&&+ \left[1-\frac{\xi}{(\zeta+\alpha)^{3}}-
\frac{2vi\zeta\sqrt{(\zeta+\alpha)^{2}-1}}{(\zeta+\alpha)
(\zeta+2\alpha)^{2}} \right]^{-1}\frac{\omega}{\Delta} \nonumber \\
&&+..., \label{21}
\end{eqnarray}
where $\alpha$ is a small constant determined by
\begin{eqnarray}
i\frac{\alpha}{\zeta+\alpha}\sqrt{(\zeta+\alpha)^{2}-1}=\frac{v\zeta}{\zeta+2\alpha}.
\label{22}
\end{eqnarray}
For sufficiently small $v$, $\alpha$ can be ignored, and
Eq.(\ref{21}) can be reduced as
\begin{eqnarray}
u=&&i\sqrt{\zeta^{2}-1}\nonumber \\
&&+\zeta^{2} \left(\zeta^{2}-1-2iv\sqrt{\zeta^{2}-1} \right)^{-1}\frac{\omega}{\Delta} \nonumber \\
&&+... \label{23}
\end{eqnarray}

In the following, in order to show what are changed by the
spin-orbit coupling, we calculate the density of states, which can
be measured by STM. The density of states is given in terms of the
Green's function by
\begin{eqnarray}
N_{S}(\omega)&=&\frac{1}{2\pi}\text{Im}\int\frac{d^{2}k}{(2\pi)^{2}}\text{Tr}[G(k,\omega)] \nonumber \\
&=&N_{0}\text{Im} \left(\frac{\tilde{\omega}}{\sqrt{\tilde{\Delta}^{2}-\tilde{\omega}^{2}}} \right) \nonumber \\
&=&N(0)\text{Im} \left(\frac{u}{\sqrt{1-u^{2}}}\right). \label{24}
\end{eqnarray}
Making use of Eq.(\ref{16}) (here we can also safely ignore the last term),
the above formula reduces to a more convenient form:
\begin{eqnarray}
N_{S}(\omega)=N(0)\zeta^{-1}\text{Im} u. \label{25}
\end{eqnarray}

First, let us consider $\zeta<1$ case, by using Eq.(\ref{19}), we
substitute the expression of $u$ into Eq.(\ref{25}) and obtain
\begin{eqnarray}
N_{S}(\omega)&=&\left\{
\begin{aligned}
         &0&   \omega<\omega_{g} \\
         &N(0)\zeta^{-\frac{2}{3}}B^{-\frac{1}{2}}\sqrt{\frac{2(\omega-\omega_{g})}{3\Delta}}&   \omega\geq\omega_{g} \\
       \end{aligned}
       \right.,
 \nonumber \\
 B&=&(1-\zeta^{\frac{2}{3}})^{\frac{1}{2}}+\frac{v(8-6\zeta^{\frac{2}{3}})}{3(2-\zeta^{\frac{2}{3}})^{3}}. \label{26}
\end{eqnarray}
From Eq.(\ref{26}), we see that $\omega_{g}$ gives the threshold
frequency of the density of states, and spin-orbit coupling
reduces the value of $N_{S}(\omega)$ for $\omega>\omega_{g}$. This
is obviously right, since from Eq.(\ref{18}), we see spin-orbit
coupling reduces the value of $\omega_{g}$. To guarantee $\int
N_{s}(\omega)d\omega$ invariant, the reduction of $N_{S}(\omega)$
is necessary.

Second, for $\zeta>1$ case, by Eq.(\ref{23}) we obtain
\begin{eqnarray}
N_{S}(\omega)&=& N(0)[(1-\zeta^{-2})^{\frac{1}{2}}+\frac{2v\zeta}{\sqrt{\zeta^{2}-1}[4v^{2}+(\zeta^{2}-1)]}\frac{\omega}{\Delta} \nonumber \\
&+&\frac{\zeta[3+2v(\zeta^{2}-5)][(\zeta^{2}-1)^{2}-4v^{2}(\zeta^{2}-1)]}{2\sqrt{\zeta^{2}-1}[(\zeta^{2}-1)^{2}+4v^{2}(\zeta^{2}-1)]^{2}}]
(\frac{\omega}{\Delta})^{2}, \nonumber \\
\label{27}
\end{eqnarray}
here we have reconsidered the second order of $\omega/\Delta$.
From Eq.(\ref{27}), we can see $N_{s}$ is finite at $\omega=0$ and
given by $N_{S}(0)=(1-\zeta^{-2})^{1/2}N(0)$. This is just a
characteristic of the gapless region, where the energy spectrum
$\omega_{g}$ vanishes though the gap parameter $\Delta$ is not
zero. This is an example of gapless superfluidity, the existence
of Cooper pairs without the existence of an energy gap in the
spectrum of excitations.

\begin{figure}
\subfigure{\includegraphics[width=9cm,height=6cm]{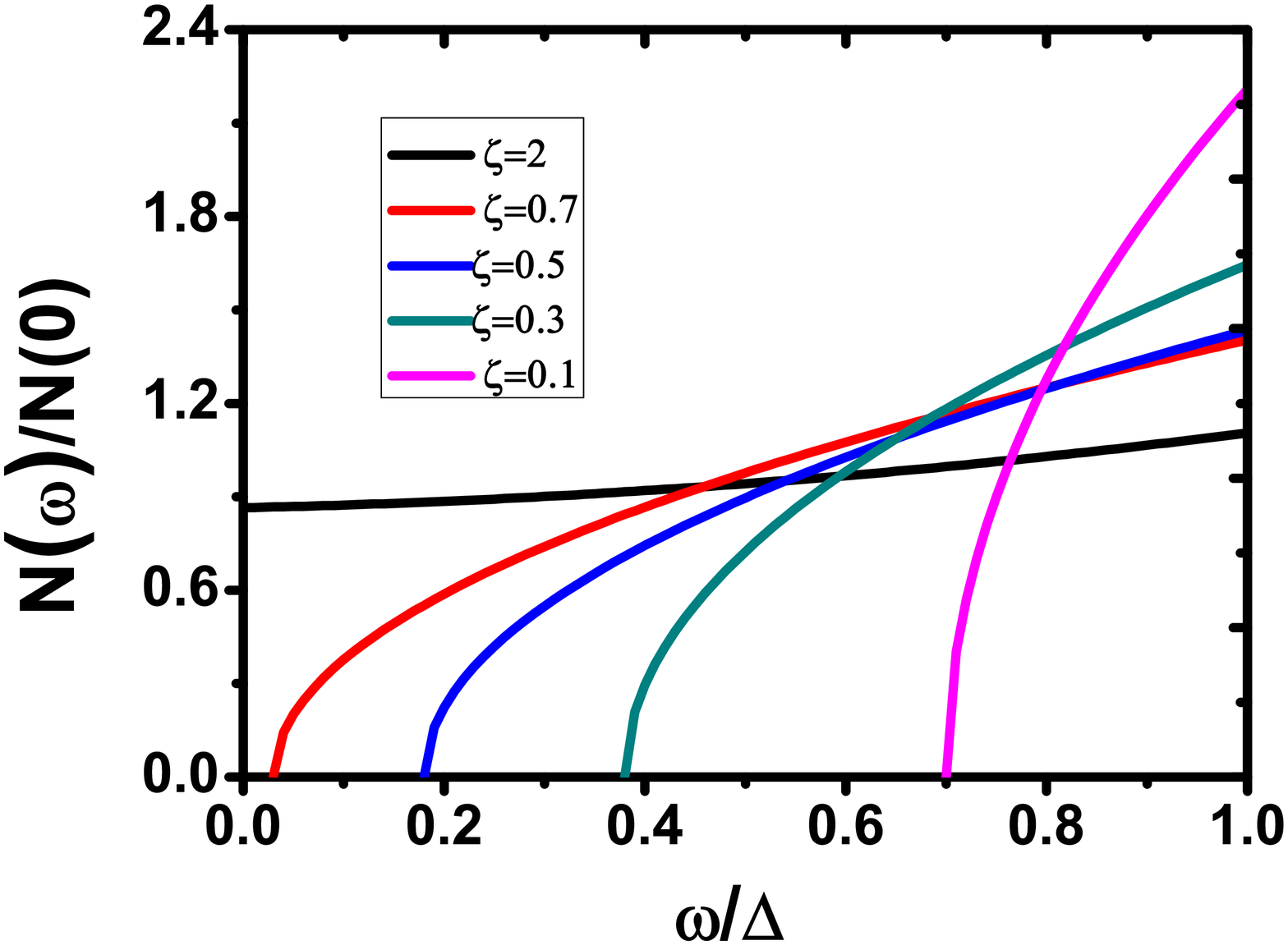}}
\caption{ (Color online) Density of states plotted for different values of $\zeta$. The parameter $v$ is set to 0.1.
} \label{Fig.3}
\end{figure}

From Eq.(\ref{27}), we also find that, for $\omega/\Delta$
sufficiently small, $N_{S}(\omega)\propto\omega$, which is quite
different from $N_{S}(\omega)\propto\omega^{2}$ in the absence of
spin-orbit coupling \cite{K. Maki}. From all of the above, it's
obviously that the spin-orbit coupling affects the effects of
magnetic impurities a lot.

\section{Conclusions}
\label{sec4}

In this paper, we have investigated the effects of magnetic
impurities in a two-dimensional superfluid Fermi gas with
spin-orbit coupling. In the presence of spin-orbit coupling, by
using the methods of t-matrix, we find that the number of bound
states has doubled comparing to the case without spin-orbit
coupling. By using the methods of Green's function, we also find
that the spin-orbit coupling changes the energy gap in the energy
spectrum and density of states. In the gap region, the larger the
concentration, the stronger the impact is given by the spin-orbit
coupling. In the gapless region (here limited to $\zeta>1$), we
find spin-orbit coupling changes the relation between $N(\omega)$
and $\omega$ from $N(\omega)\propto\omega^{2}$ to
$N(\omega)\propto\omega$ in the small $\omega$ limit.

Since we only consider the spin-flip effects of the magnetic
impurities and totally ignore the spin-exchange effects, the
impurities are classical impurities and as a result, the Kondo
effect is absent. In future, such a system including spin-exchange
term may be used to explore the effects of spin-orbit coupling to
the Kondo problem in cold atomic systems. In addition, the effects
of magnetic impurities in imbalance Fermi gas with spin-orbit
coupling are also worth exploring.

\begin{acknowledgments}
We thank Liang Chen for helpful discussions. This work is supported by NSFC Grant No.10675108.
\end{acknowledgments}

\appendix
\section{Appendix}

To obtain Eq.(\ref{7}), we first write down the Green's function in the absence of a magnetic impurity,
\begin{eqnarray}
G_{0}(k,\omega)=\frac{1}{\omega-\varepsilon_{k}\tau_{3}+\Delta\rho_{1}\tau_{3}-\alpha_{R}k(e^{i\varphi_{k}}\tau_{+}
+e^{-i\varphi_{k}}\tau_{-})}, \nonumber \\
\end{eqnarray}
$G_{0}(k,\omega)$ is a $4\times4$ matrix, after we write down its entities, by using the formula
\begin{eqnarray}
F(\omega)=\sum_{k}G_{0}(k,\omega),
\end{eqnarray}
we can integrate out $k$ directly, and obtain the entities of
$F(\omega)$. However, because the presence of spin-orbit coupling,
the energy spectrum has greatly modified, and this makes the
integration difficult. To avoid this difficulty, we assume the
strength of spin-orbit coupling, $\alpha_{R}<<V_{F}$, where
$V_{F}=k_{F}/m$ is the Fermi velocity. Following the assumption
$k-k_{F}\approx m\epsilon/k_{F}$ made by G. Rickayzen when he
solved the problem of impurities in metal\cite{G. Rickayzen},
assuming the particle-hole symmetry of Fermi band and
approximating the normal-state density of states by the value
$\rho(0)$ at the Fermi level, we obtain
\begin{eqnarray}
F(\omega)=-\pi\rho(0)\frac{\omega-\Delta\rho_{1}\tau_{3}-m\alpha_{R}^{2}\rho_{3}}{\sqrt{\Delta^{2}-\omega^{2}}}.
\end{eqnarray}
Using this formula and Eq.(\ref{5}), we obtain the equation for the bound-state energies as
\begin{eqnarray}
0=\text{det}[1-AF(\omega)^{2}],
\end{eqnarray}
where $A=(Us\pi\rho(0)/2)^{2}$. By directly solving a eigenvalue
problem and assuming $m\alpha_{R}^{2}/\Delta<<1$, we obtain
\begin{eqnarray}
\omega_{1}&=&\pm(1-\frac{1}{8}A\frac{m^{4}\alpha_{R}^{8}}{\Delta^{4}})\Delta, \nonumber \\
\omega_{2}&=&\pm(\frac{1-A}{1+A}-\frac{A}{1+A}\frac{m^{2}\alpha_{R}^{4}}{\Delta^{2}})\Delta.
\end{eqnarray}


\begin{thebibliography}{99}
\bibitem{M. Z. Hasan} M. Z. Hasan, C. L. Kane, Rev. Mod. Phys. {\bf 82}, 3045 (2010).
\bibitem{X. L. Qi} X. L. Qi, S. C. Zhang, Rev. Mod. Phys. {\bf 83}, 1057 (2011)
\bibitem{C. A. Regal} C. A. Regal, M. Greiner, and D. S. Jin, Phys. Rev. Lett. {\bf 92}, 040403 (2004)
\bibitem{M. Bartenstein} M. Bartenstein, A. Altmeyer, S. Riedl, S. Jochim, C. Chin, J. H. Denschlag,
 and R. Grimm, Phys. Rev. Lett. {\bf 92}, 120401 (2004)
\bibitem{M. W. Zwierlein} M. W. Zwierlein, C. A. Stan, C. H. Schunck, S. M. F. Raupach, A. J. Kerman,
and W. Ketterle, Phys. Rev. Lett. {\bf 92}, 120403 (2004)
\bibitem{J. Kinast} J. Kinast, S. L. Hemmer, M. E. Gehm, A. Turlpov, and J. E. Thomas, Phys. Rev. Lett. {\bf 92}, 150402 (2004)
\bibitem{Y. -J. Lin1} Y. -J. Lin, R. L. Compton, A. R. Perry, W. D. Phillips, J. V. Porto, and I. B. Spielman, phys. Rev. Lett. {\bf 102}, 130401 (2009)
\bibitem{Y. -J. Lin2} Y. -J. Lin, K.Jimenez-Garcia, and I. B. Spielman, Nature {\bf 471}, 83 (2011).
\bibitem{Xiaosen Yang} Xiaosen Yang, Shaolong Wan, Phys. Rev. A {\bf 85}, 023633 (2012).
\bibitem{Rok Zitko} Rok Zitko and Janez Bonca, Phys. Rev. B {\bf 84}, 193411 (2011).
\bibitem{Mahdi Zarea} Mahdi Zarea, Sergio E. Ulloa, and Nancy Sandler, Phys. Rev. Lett. {\bf 108}, 046601 (2012)
\bibitem{L. Isaev} L. Isaev, D. F. Agterberg, and I. Vekhter, Phys. Rev. B {\bf 85}, 081107 (2012)
\bibitem{A. A. Abrikosov} A. A. Abrikosov and L. P. Gor'kov, Sov. Phys.¡ªJETP {\bf 15}, 752(1962).
\bibitem{K. Maki} K. Maki, in $Superconductivity$, edited by R. D. Parks (Marcel Dekker, New York, 1969),p1045-1046.
\bibitem{C. Zipkes} C. Zipkes, S. Palzer, C. Sias, and M. K¡§ohl, Nature (London) {\bf 464}, 388 (2010).
\bibitem{G. Modugno} G. Modugno, Rep. Prog. Phys. {\bf 73}, 102401 (2010).
\bibitem{A. Schirotzek} A. Schirotzek, C. H. Wu, A. Sommer, and M. W. Zwierlein, Phys. Rev. Lett. {\bf 102}, 230402 (2009).
\bibitem{S. Nascimbene}  S. Nascimbene, N. Navon, K. J. Jiang, L. Tarruell, M. Teichmann, J. McKeever, F. Chevy, and C. Salomon, Phys. Rev. Lett. {\bf 103}, 170402 (2009).
\bibitem{Y. Ohashi} Y. Ohashi, Phys. Rev. A {\bf 83},063611 (2011).
\bibitem{H. Shiba} H. Shiba, Prog. Theor. Phys. {\bf 40}, 435 (1968).
\bibitem{Alexander Altland} Alexander Altland and Ben Simons, $Condensed$ $Matter$ $Field$ $Theory$, Chap.6. Cambridge Univ. Press (2006).
\bibitem{Lei Jiang} Lei Jiang, Leslie O. Baksmaty, Hui Hu, Yan Chen, and Han pu, Phys. Rev. A {\bf 83} 061604 (2011)
\bibitem{G. Rickayzen} G. Rickayzen, $Green's$ $Functions$ $and$ $Condensed$ $Matter$, Chap.4. Academic Press (1980).

\end{thebibliography}
\end{document}